\documentclass[12pt]{article}
\newcommand{\beq}{\begin{equation}}
\newcommand{\eeq}{\end{equation}}

\usepackage{color,
cite,graphicx}

\newcommand{\rf}[1]{(\ref{#1})}

\newcommand{\m}{\mu}
\newcommand{\ep}{\varepsilon}

\begin{document}
\topmargin 0pt
\oddsidemargin 5mm
\headheight 0pt
\headsep 0pt
\topskip 5mm

\begin{center}
\hspace{10cm}
 
\vspace{48pt}
{\large \bf
A REINTRODUCTION OF DYNAMICAL SU(2) GAUGE FIELDS IN HUBBARD MODELS }
\end{center}

\renewcommand{\thepage}{{\roman{page}}}

\vspace{40pt}
 
\begin{center}
{\bf Peter Orland}\footnote{orland@nbi.dk}
 
\vspace{30pt}
 
The Niels Bohr Institute, \\
Blegdamsvej 17, DK-2100, \\
Copenhagen {\O}, Denmark, \\

\vspace{15pt}
 
Physics Department, \\
Baruch College, \\ The City University of New York, \\
17 Lexington Avenue, \\
New York, NY 10010, U.S.A.,\\

\vspace{15pt}

and\\

\vspace{15pt}

Physics Ph.D. Program,\\The Graduate School\\ and University Center,\\ The City University of New York,
\\ 365 Fifth Avenue,\\
New York, NY 10016, U.S.A.

\end{center}
 
\vfill
 
\begin{center}
{\bf Abstract}
\end{center}
This is a brief discussion of an old preprint (which follows). This 
paper explains how non-Abelian gauge magnets 
originate as effective dynamics in models of 
hopping particles. In particular, an explicit model is discussed in which both link and plaquette terms appear. The motivation to reintroduce the idea is some recent theoretical progress on the topic of optical lattices.

\vfill

\newpage

\section*{An Apology}

Recently there have appeared some intriguing results on how dynamical gauge invariance may occur in
optical lattices \cite{TagCelZamLew}. In particular, it appears that Abelian gauge magnets of the type
discussed in \cite{AbelianNuclPhys} could arise. The authors of Reference \cite{TagCelZamLew} present a detailed discussion as to how such a model can be simulated. In the light of these developments,  it may be timely to reintroduce the paper from 1990, entitled ``SU(2) Gauge Invariance in Hubbard Models and Superconductivity". The main point was that a non-Abelian gauge magnet appears in the hopping-parameter expansion of a particular Hubbard model. The paper has been available only as a scanned manuscript \cite{KEKlink} until now.

The paper concerns only dynamical non-Abelian gauge fields. Background non-Abelian gauge fields have been discussed in References \cite{background}.

I have not revised the paper except to update the references and to correct a few misprints. The model was proposed to explain 
copper-oxide-layer superconductivity, through either confinement or screening of spin. Applying the model to
optical lattices may be worthy of investigation.

The only other reason for this apology is to mention that there is a very general context in which gauge magnets \cite{gm} (also known as ``quantum link models"), both Abelian and non-Abelian, should appear at low frequencies. Indeed, there appears to be a general theorem concerning how such models arise in the hopping-parameter expansion. I intend to explain how this theorem works elsewhere.

\newpage

\setcounter{page}{1}

\renewcommand{\thepage}{{\arabic{page}}}

 
\begin{flushright}
NBI-HE-90-29\\
\hfill
June 1990
\end{flushright}

\begin{center}
\hspace{10cm}
 
\vspace{48pt}
{\large \bf
SU(2) GAUGE INVARIANCE IN HUBBARD MODELS AND SUPERCONDUCTIVITY }
\end{center}

\vspace{40pt}
 
\begin{center}
{\bf Peter Orland}
 
\vspace{30pt}
 
The Niels Bohr Institute, \\
Blegdamsvej 17, DK-2100, \\
Copenhagen {\O}, Denmark \\

\vspace{15pt}

and\\

\vspace{15pt}
 
Physics Department, \\
Baruch College, \\ The City University of New York, \\
17 Lexington Avenue, \\
New York, NY 10010, U.S.A.
\footnote{Permanent address after Sept.1, 1990}\\

\end{center}
 
\vfill
 
\begin{center}
{\bf Abstract}
\end{center}
It is suggested that in doped copper oxide layers, the lowest energy $p_{x}$
oxygen orbital for a hole is split by lattice distortions, into states which
hybridize asymetrically with the $d_{x^2-y^2}$ orbitals on each of the
neighboring copper atoms. The appropriate Hubbard model has two available sites
associated with each oxygen atom. The system is effectively
described by an ${\rm SU}(2)$ gauge theory, with an additional coupling
to a charged spinning superfluid. Spin is thereby either confined
or screened. Both  possibilities lead to hole pairing and
superconductivity.

\vfill

\newpage
 
\section{Introduction}
 
Magnetic, rather than phononic, dynamics is widely believed to be
responsible for high temperature superconductivity \cite{all}, \cite{Wie}.
Most of these ideas are centered around the notion that holes
in metal oxide layers are described by an effective one-band Hubbard model
resulting from integrating out degrees of freedom on the oxygen sites
\cite{Zhang}, or copper sites \cite{Em}. The bond
between an oxygen atom and the
two neighboring copper atoms is assumed to be a sigma bond in which the $p_{y}$
ground state orbital on the oxygen atom atoms hybridizes strongly
with both $d_{x^2-y^2}$ orbitals on the copper atoms.
There has been much speculation as to the nature of the
the ground state of this model, much of it involving novel physical
and mathematical ideas \cite{all}, \cite{Wie} in particular
anyons \cite{Laughlin}. Here a different starting point is suggested.
 
If the effect of oxygen nuclear motion is included, the $p_{y}$ orbital
becomes two states. Consider the situation depicted in fig.1abc. If
the hole in this orbital is closer to one of the two copper
atoms, say, atom $A$ the oxygen atom will be pushed slightly towards or
away from the other copper atom, called atom $B$. The sign of the pushing
depends upon Coulombic as well as collective effects. It is assumed
here that the overall consequence is that
hybridization with the $d$ orbital on atom $A$ strengthens, while
hybridization with the $d$ orbital on atom $B$ weakens. The effect of the
local distortions of the lattice is that there is a double-well potential
which must be added to the atomic potential in the Hamiltonian. The
resulting Hubbard model has two sites instead of one associated with
each oxygen atom. The ground state oxgen orbital is still $p_{y}$, but
there is now an excited state whose wave function is symmetric
along the x-axis.
 
It is important to stress that this proposal is not a
B.C.S. picture. The lattice distortions have wavelengths the size of the
interatomic spacing
and do not give rise to long range forces by themselves. The mechanism
of superconductivity is essentially magnetic.
 
A second assumption is also made; it is that
the holes in the vicinity of a copper atom tend to form a spin singlet.
The result is, after integrating out high frequency modes, an ${\rm SU}(2)$
lattice gauge theory, of the type studied in \cite{Horn}, \cite{O+R}. The
full gauge group is ${\rm SU}(2) \times {\rm U}(1)$, including
electromagnetism. The
calculation is done perturbatively, much like that done for the one-band
Hubbard model at half-filling to obtain the Heisenberg model (Mott-Hubbard
insulator) \cite{Ian}. These Hamiltonian lattice gauge theories, named gauge
magnets in ref. \cite{O+R}, are rather different in structure from conventional
lattice gauge theories. In particular, they are formulated in terms
of only one representation of the gauge group. An ${\rm SU}(2)$ gauge transformation
is the total spin in the vicinity of a copper atom; the two ``colors" of the
gauge theory are just $\uparrow $ and $\downarrow $. This is closly related to the
nondynamical gauge invariance noted by Baskaran and Anderson in the usual
half-filled Hubbard model \cite{A}.
 
The effective gauge theory obtained has a superfluid, or Higgs field. It
is in either the confined or the Higgs
phase. General arguments \cite{FS}
imply that that in either case the cell excitations are
tightly paired into ``baryon" excitations analogous to those in Q.C.D.
Since the gauge group is ${\rm SU}(2)$ (instead of the ${\rm SU}(3)$ color group in
Q.C.D.) these excitations are bound states of two cell excitations. The
picture has some similiarities with the ${\rm U}(1)$ gauge theory
confinement schemes discussed by Wiegmann \cite{Wie} and
Fradkin and Kivelson \cite{FK}.
The Higgs field is (fractionally) charged, and so is a second (non-Cooper
pair) carrier of supercurrent.
 
The situation on an oxygen atom
is described (before considering hybridization) by the two-site
Hamiltonian:
\beq
H=t_{O}\; \sum_{\alpha }\;
c^{\dag} _{\; \vec i ,\alpha }c_{\vec j ,\alpha }\; ,   \label{10.1}
\eeq
where $\vec i $, and $\vec j $ are the different site locations,
$\alpha = \uparrow , \downarrow $ and $c^{\dag} _{\; \vec i ,\alpha },c_{\; \vec i ,\alpha }$
are the creation and annihilation operators for holes. This Hamiltonian
has a symmetric excited state lying at an energy twice the (wrong sign)
oxygen hopping parameter, $t_{O}$, above the antisymmetric ground state.
 
A more precise statement of the second assumption is that, if one ignores
the hopping between the two oxygen sites \rf{10.1}, then the ground state
of the hole configuration on a copper site and the adjacent oxygen
sites is a spin singlet, while the (spin degenerate) first excited state is
not. The {\em cell \/} in the vicinity of the copper atom will be defined
to be this set of sites. Spin non-singlets are excitations which can
move from cell to cell through the lattice. The essential point of this paper
is that they must move in a gauge covariant manner. The Gauss's law operator
$\vec G$ is the sum of two terms. The first
term is the total spin $\vec S$ in a particular cell. The second term is minus
the sum over first excited states $|X>$ of the excited state spin eigenvalue
$\vec S_{X} $ times the projection operator $|X>\; <X|$ for that particular
excited state. This Gauss's law operator
will, by construction, obey the appropriate local commutation relations
and annihilate  physical states.
By definition then Gauss's law is satisfied. Therefore ${\rm SU}(2)$ gauge
invariance of the states is inevitable.
 
There are some possible objections that might be raised to the ideas
presented here. The fact that motion of oxygen atoms is essential
seems to suggest that a charged density wave would form. This is not
true, as it is inconsistent with gauge invariance.
Another objection might be that there are four states, rather than
two on the oxygen orbitals. This is in fact so, but these states are
not degenerate, and the ground state orbital is not significantly
different from that indicated by experiments.
 
This article is a slightly revised version of a paper circulated
in March, 1990, while the author was at Virginia Polytechnic Institute 
and State University.

\section{Gauge Magnets}
 
Lattice gauge magnets \cite{Horn},\cite{O+R} are gauge invariant
generalizations of isotropic Heisenberg magnets. They are formulated
quite differently from the usual Wilson or Kogut-Susskind lattice
gauge theories. The first ${\rm SU}(2)$
gauge magnet Hamiltonian was written down by Horn \cite{Horn}
who proposed it as a simple regularization of
Yang-Mills theory. The author and D. Rohrlich \cite{O+R} showed
that the Horn model has a nonrelativistic spin wave dispersion relation. It
was also found that there is an enormous variety of ${\rm SU}(2)$
gauge magnets. Abelian gauge magnets have been studied as way of formulating
short range resonating valence bond phases \cite{FK}, \cite{SRRVB}. For a
more detailed discussion, see ref. \cite{O+R}.
 
In order to define gauge magnets for the spin-1/2 representation
of ${\rm SU}(2)$, it is necessary to consider operators at the links of a square
lattice, ${\vec x}, \hat m$ connecting the sites ${\vec x}$ and ${\vec x} +a\hat m,
\; m=1,..,d$, where ${\vec x}$ is a d-component site vector. The Hilbert space
at each link is four-dimensional, so these operators can be thought of
four-by-four matrices acting on a given link. These operators are
Dirac matrices for a Euclidean metric, $\gamma^{0}({\vec x},m),
\; \gamma^{1}({\vec x},m),\;
\gamma^{2}({\vec x},m),\; \gamma^{3}({\vec x},m) $, with the anticommutation relations
\beq
[\gamma^{\mu}({\vec x},m),\gamma^{\nu}({\vec x},m)]_{+} = \delta ^{\mu \nu}, \label{1.1}
\eeq
on the same link and the commutation relations
\beq
[\gamma^{\mu}({\vec x},m),\gamma^{\nu}(\vec y,n)] = 0, \label{1.2}
\eeq
on different links. It is important to emphasize that the greek indices $\mu,
\nu$ simply label different operators, and have nothing to do with space or
time. The index $m$ was called $i$ in ref. \cite{O+R}. A specific representation
at one link is
\beq
\gamma^{0}=\left( \begin{array}{cc}
         0 & 1 \\
         1 & 0 \\  \end{array} \right) \; , \;\;
\vec \gamma =\left( \begin{array}{cc}
                 0 & i\vec \sigma \\
        -i\vec \sigma & 0 \\  \end{array} \right),     \label{1.3}
\eeq
where $\vec \sigma $ are the usual Pauli matrices.
Other useful operators are
\beq
\gamma^{5}=\gamma^{0}\gamma^{1}\gamma^{2}\gamma^{3}\; ,\; \rho ^{\mu}=-i\gamma^{5}\gamma^{\mu}\;, \;
\sigma ^{\mu \nu}=-\frac{i}{4} [\gamma^{\mu},\gamma^{\nu}],             \label{1.4}
\eeq
and
\beq
\Sigma ^{a} = \frac{1}{2} \sum_{bc} \varepsilon ^{a b c} \sigma ^{bc}
-\sigma ^{0a},\;\;
{\tilde{\Sigma}} ^{a} = \frac{1}{2} \sum_{bc} \varepsilon ^{a b c} \sigma ^{bc}
+\sigma ^{0a},\;\; a,b,c=1,2,3.                                       \label{1.5}
\eeq
In the representation \rf{1.3}
\beq
\vec \Sigma =\frac{1}{2} \left( \begin{array}{cc}
         \vec \sigma & 0 \\
                0 & 0 \\  \end{array} \right) \; , \;\;
\vec{{\tilde{\Sigma}}} =\frac{1}{2} \left( \begin{array}{cc}
                 0 & 0 \\
                 0 & \vec \sigma \\  \end{array} \right).     \label{1.6}
\eeq
While the introduction of these operators may seem rather {\em ad hoc\/} at
this stage, it will be shown in the next section that they can arise naturally
in a particular Hubbard model of holes.
 
The basic lattice gauge fields are
\beq
U_{\alpha \beta}({\vec x} ,m)=\gamma^{0}-i{\vec \gamma} \cdot{\vec x} , \;,    \label{1.7}
\eeq
and
\beq
U^{5} _{\alpha \beta}({\vec x} ,m)=\rho ^{0}({\vec x} ,m)\; \delta_{\alpha \; \beta}
-i\vec \rho ({\vec x} ,m) \; \cdot \; \vec \tau_{\alpha \; \beta} \;.    \label{1.8}
\eeq
Here the $2 \times 2$ matrices $\tau ^{1},\tau ^{2},\tau ^{3}$ are again
the Pauli matrices. The gauge fields $U({\vec x},m)$ and $U^{5}({\vec x},m)$
should be thought of as operator
valued $2 \times 2$ matrices; the indices $\alpha, \beta$ in \rf{1.7} and
\rf{1.8} are simply labels of matrix rows and columns. The matrices
$\tau ^{a}$ do not act on the Hilbert space. The
operators $\sigma ^{a}$ {\em do} act on the upper two components or lower
two components of the Hilbert space, however.
 
The ``vacuum" generators of gauge transformations are
\beq
\vec G ({\vec x})=\sum_{m} [ \vec{{\tilde{\Sigma}}} ({\vec x} ,m)
+\vec \Sigma ({{\vec x}} - a {\hat m} ,m) ] \;  \label{1.9}
\eeq
and obey the local commutation relations
\beq
[ G^{a} ({\vec x} ), G^{b} (\vec y ) ]= 2i\sum_{c}
\ep ^{abc} \delta_{{\vec x} \vec y}
G^{c}({\vec x}).         \label{1.10}
\eeq
The fields $U, U^{5}$ transform as ``parallel transport" or ``connection" fields:
If $Y({\vec x} ,m)$ is any linear combination of $U({\vec x} ,m)$
and $U^{5}({\vec x} ,m)$ (with complex coefficients which can depend
on the link) then :
\beq
[ G^{a}({\vec x}), Y({\vec x} ,m)]=-i\tau ^{a}
Y({\vec x} ,m) \; ,\;\;
[ G^{a}({\vec x}), Y({\vec x} - a\hat m ,m)]=i Y({\vec x} ,m)
\tau ^{a}    \;.        \label{1.13}
\eeq
The right-hand-sides in \rf{1.13} are matrix products over greek indices.
It is simple to make gauge invariant quantities by multiplying $U$'s
together, end to end. Some examples considered in ref. \cite{O+R}
were the gauge magnet
Hamiltonians:
\beq
H=J\sum_{{\vec x} } \sum_{ m \neq n}\; {\rm Tr}
U({\vec x} ,m)U({\vec x} +\hat m,n)U({\vec x} +\hat n,m)^{\dagger}
U({\vec x} ,n)^{\dagger}  \;,           \label{1.14}
\eeq
(the trace is over greek indices)
which has a nonrelativistic spin wave dispersion relation,
and, in two spatial dimensions, the ``staggered" model:
\begin{eqnarray}
H &=& J\sum_{x^{1}+x^{2} even} \; {\rm Tr}
      U({\vec x} ,1)U({\vec x} +\hat 1 ,2)U({\vec x} +\hat 2,1)^{\dagger}
      U({\vec x} ,2)^{\dagger}  \;,   \nonumber \\
  &+& K\sum_{x^{1}+x^{2} odd}
\; {\rm Tr} U^{5}({\vec x} ,1)U^{5} ({\vec x} +\hat 1 ,2)
U^{5} ({\vec x} +\hat 2,1)^{\dagger}
U^{5} ({\vec x} ,2)^{\dagger}  \;,   \label{1.15}
\end{eqnarray}
which has a relativistic massive dispersion relation, and was argued to
be topologically massive ${\rm SU}(2)$ Yang-Mills theory \cite{Jackiw} in
ref. \cite{O+R}.
Another operator which commutes with all the $G^{a}$'s
is $\gamma^{5}({\vec x} ,m)$, so this is also a possible term to include in
a gauge magnet Hamiltonian.
 
Coupling a fermionic matter field $c_{{\vec x} , \alpha}$ to the gauge field
\beq
Y({\vec x} ,m) = r({\vec x} ,m)U({\vec x} ,m)
+r^{5} ({\vec x} ,m)U^{5}({\vec x} ,m)       \label{1.151}
\eeq
is accomplished with
\beq
H^{1}_{eff}=-T\sum_{{\vec x} ,m} \sum_{\alpha ,\beta}
c^{\dag}_{{\vec x} ,\alpha} \;
Y_{\alpha \; \beta}({\vec x} ,m) \; c_{{\vec x} ,\beta} \; +\; h.c. , \label{1.16}
\eeq
with the Gauss's law operator modified to
\beq
\vec G ({\vec x})=\sum_{m} [ \vec{{\tilde{\Sigma}}} ({\vec x} ,m)
+\vec \Sigma ({\vec x} -a\hat m ,m) ] \; +\;
\frac{1}{2} \sum_{\alpha \beta }
c^{\dag} _{{\vec x} ,\alpha }
\vec{\tau}_{\alpha \beta}
c _{{\vec x} ,\alpha } \; .                 \label{1.17}
\eeq
 
\section{The Hopping Parameter Expansion}
 
A model Hubbard Hamiltonian with the features discussed in the
introduction will now be
studied in perturbation theory. It is somewhat unrealistic as the
doping is far too large, and there is a hole for every oxygen atom.
The basic idea should extend, however, to the case of a more realistic
doping concentration, as will be discussed at the end
of this section. Perturbation theory is not quantitatively correct
unless all hopping parameters are small. Nonetheless, it should be
a good guide to the form of the effective Hamiltonian.
 
This two-dimensional Hubbard model describes the dynamics of holes hopping
between sites $\vec i$ on
the lattice shown in fig.2. There are two available sites on each link (oxygen
atom) and one available site at each intersection point (copper atom). Each
copper atom together with the nearest neighbor sites on the adjacent oxygen
atoms is regarded as a cell. Thus each cell has five sites. We can label the
copper atoms by vectors $\vec i ={\vec x} $ and the sites on the oxygen
atoms by
$\vec i= {\vec x} \pm b\hat m $, where $m=1,2$ and $b<a/2$ is the spacing
between a copper site and the nearest oxygen site. The cell containing
${\vec x} $ and ${\vec x} \pm b\hat m$ will be denoted by $B_{{\vec x}}$. The
sites in $B_{{\vec x}}$ can be written
alternatively as $\vec i \; \ep \; B_{{\vec x}}$. The oxygen atom connecting
the sites ${\vec x}$ and ${\vec x} +a\hat m$ will be denoted by
$L_{{\vec x} ,\hat m}$.
The sites on ${\vec x} +b \hat m$ and ${\vec x} +(a-b) \hat m$
can be written as $\vec i \; \ep \; L_{{\vec x} ,\hat m}$. It is convenient to
drop the subscripts from $B$ and $L$.
 
The Hamiltonian has the form
\beq
H=H_{0}+V .                                           \label{2.1}
\eeq
The unperturbed part of \rf{2.1} is:
\begin{eqnarray}
H_{0}= U\sum_{{\vec x}}\;n_{{\vec x},\uparrow}n_{{\vec x},\downarrow}
+ J\sum_{B} \sum_{\vec i \neq \vec j \ep B}\;
\vec S_{\vec i}\; \cdot \; \vec S_{\vec j}
+A\sum_{{\vec x}}\; (n_{{\vec x}, \uparrow} + n_{{\vec x}, \downarrow})  \nonumber \\
+ D\sum_{L}\sum_{\vec i\neq \vec j \ep L}\;
(n_{\vec i,\uparrow}+n_{\vec i,\downarrow})
(n_{\vec j,\uparrow}+n_{\vec j,\downarrow})\;
+\mu [\sum_{i} (n_{\vec i,\uparrow}+n_{\vec i,\downarrow}) -h] .               \label{2.2}
\end{eqnarray}
The coefficients $U$, $J$, $A$ and $D$ are positive, and $U,D\gamma J\gamma A$.
The first term of \rf{2.2} is a repulsive interaction on copper atoms.
The second term in \rf{2.2} is an antiferromagnetic interaction between
any two holes in the cell. The third term of \rf{2.2} favors
occupation of sites on the oxygen atoms over occupation of sites at the
copper atoms. The fourth term in \rf{2.2} discourages the occupation
of any pair of sites on an oxygen atom by more than one hole. The
last term enforces the hole number to be fixed to $h$.
 
Consider the situation in which the total number
of holes, $h$ in the model is set to be $2N$, where $N$ is the total number
of copper atoms. For large $U$, $A$ and $D$ the (highly degenerate)
ground state of $H_{0}$ has two holes per cell, occupying two
different links (fig.3). The configurations resemble those
of two-dimensional cubic ice crystals (the six-vertex
model). The lowest lying excited states can be made by
taking a hole from one cell and placing it at the copper atom
in another cell. The latter cell now contains a total
of three holes, one hole at the copper atom and the remaining
two hole on two different oxygen atoms. The energy of such states
is of order $J$.
 
If the total number of holes is $(2+\epsilon )N$ then a fraction
$\epsilon $ of the cells will be excited. The lowest lying states will
have holes at some copper atoms (fig.5). In these states, excited
cells are the only cells which are not spin singlets. Their energy
is of order $\epsilon A$.
 
The interaction of \rf{2.1} introduces hopping between the sites:
\beq
V=-\sum_{<\vec i ,\vec j >} \sum_{\alpha }\; t_{\vec i ,\vec j } \;
c^{\dag} _{\vec i ,\alpha }c_{\vec j ,\alpha }\; ,                    \label{2.3}
\eeq
where the hopping parameters $t_{\vec i ,\vec j }$ are regarded as small
compared to
the constants $U, J, A$ and $D$. The hopping parameter
between two sites on the same oxygen atom will be
denoted by $t_{{\vec x} + b\hat m ,\; {\vec x} +(a-b)\hat m}=-t_{O}$, as before,
while that between copper
and oxygen sites will be denoted by $t_{{\vec x} ,\; {\vec x} \pm b\hat m}
=t_{Cu-O}$.
 
This toy model now has the basic features discussed in the introduction. The
lattice is broken up into cells, spin singlets are energetically
favorable in the cells, and there is weak hopping between the
cells. It will be verified in this section that, in the hopping parameter
expansion, with the number of holes equal to $(2+\epsilon )N$, this
system is a gauge magnet. This expansion is not quantitatively
correct, because in $CuO_{2}$ layers the parameter $t_{O}$ must be
actually bigger than $A$. In the limit that $t_{O}$ becomes
infinite, the system becomes a one-band Hubbard
model \cite{all}. What this means is that as $t_{O}/A$
increases, there is eventually a transition to a phase described by
the one-band model. Above
this phase transition, the low frequency behaviour is that of
the $t-J$ model \cite{Zhang}. It
is a crucial assumption that $t_{O}/A$
is fairly large (of one order of magnitude, say)
at the phase transition. As long as $t_{O}/A$ is below the
transition point, the form of the resulting effective Hamiltonian obtained in
perturbation theory will be correct. A pictorial comparison of the various
constants in the model are shown in fig.5.

For $\epsilon \ll 1$ most of the excited cells (with a
hole on the copper $d$ orbital) will be surrounded by
cells which are not excited. It is straightforward to see how
an excitation moves through the lattice. Since $A$ is the smallest
of the constants in \rf{2.1} the most significant energy
denominator is $1/A$.
 
Consider the configurations of two adjacent cells in fig.6ab. The cell on the
left is excited, while that on the right is not. The holes at sites other
than the two copper atoms and the oxygen atom joining the cells are
superfluous, so the configurations of fig.6a and fig.6b are conveniently
labeled by the spin at these four sites, on a line from left to right:
\beq
|\uparrow ,0,\uparrow ,0>,\; |\uparrow ,0,\downarrow ,0>,\;
|\downarrow ,0,\uparrow ,0>,\; |\downarrow ,0,\downarrow ,0>\; ,      \label{2.7}
\eeq
and
\beq
|\uparrow ,\uparrow ,0,0>,\; |\uparrow ,\downarrow ,0,0>,\;
|\downarrow ,\uparrow ,0,0>,\; |\downarrow ,\downarrow ,0,0>\; ,      \label{2.8}
\eeq
respectively. The left-most spin is at ${\vec x}$, while the right-most
spin is at ${\vec x} + a\hat m$.
 
The states \rf{2.7} can undergo the following changes under hopping:
\begin{eqnarray}
|\uparrow ,0,\uparrow ,0> &\rightarrow & |\uparrow ,0,0,\uparrow >\;\; \rightarrow \;\;
|0,\uparrow ,0,\uparrow >                 \nonumber \\
|\uparrow ,0,\downarrow ,0> &\rightarrow & |\uparrow ,0,0,\downarrow > \;\; \rightarrow  \;\;
|0,\uparrow ,0,\downarrow >                 \nonumber \\
|\downarrow ,0,\uparrow ,0> &\rightarrow & |\downarrow ,0,0,\uparrow > \;\; \rightarrow  \;\;
|0,\downarrow ,0,\uparrow >                 \nonumber \\
|\downarrow ,0,\downarrow ,0> &\rightarrow & |\downarrow ,0,0,\downarrow > \;\; \rightarrow  \;\;
|0,\downarrow ,0,\downarrow >                                              \label{2.9}
\end{eqnarray}
The intermediate states are short lived; the lifetime is of order
$\hbar /A$. The matrix elements of the Hamiltonian between the initial
and final states of \rf{2.9} are to second order in perturbation theory
given by
\begin{eqnarray}
<\uparrow ,0,\uparrow ,0|\; H\; |0,\uparrow ,0,\uparrow >
&=&<\uparrow ,0,\uparrow ,0|\; H\; |0,\uparrow ,0,\uparrow >
\; =\; <\uparrow ,0,\uparrow ,0|\; H\; |0,\uparrow ,0,\uparrow >    \nonumber \\
&=&<\uparrow ,0,\uparrow ,0|\; H\; |0,\uparrow ,0,\uparrow >
\; =\; -t_{Cu-O} ^{2}/A \; .                                   \label{2.10}
\end{eqnarray}
It is now possible to write an effective Hamiltonian for the low lying
states; these do not include the intermediate states of \rf{2.10}. This
amounts to integrating out modes of frequency $A/\hbar $, while ignoring
modes of higher frequency.
Consider the operators $Y^{0}({\vec x} ,m),\vec Y ({\vec x} ,m)$ acting only
on the on the spins of the oxygen atom by
\begin{eqnarray}
Y^{0}({\vec x} ,m) &=& \sum_{\alpha}
c^{\dag}_{{\vec x} +b \hat m ,\alpha}
c_{{\vec x} +(a-b) \hat m ,\alpha} \; ,       \nonumber  \\
Y^{1}({\vec x} ,m) &=& i\sum_{\alpha}
c^{\dag}_{{\vec x} +b \hat m ,\alpha}
c_{{\vec x} +(a-b) \hat m ,-\alpha} \; ,       \nonumber  \\
Y^{2}({\vec x} ,m) &=& -\sum_{\alpha}
\; sgn(\alpha)\; c^{\dag}_{{\vec x} +b \hat m ,\alpha}
c_{{\vec x} +(a-b) \hat m ,-\alpha} \; ,       \nonumber  \\
Y^{3}({\vec x} ,m) &=& i\sum_{\alpha}
\; sgn(\alpha)\; c^{\dag}_{{\vec x} +b \hat m ,\alpha}
c_{{\vec x} +(a-b) \hat m ,\alpha}  \; ,      \label{2.11}
\end{eqnarray}
with the conventions $sgn(\uparrow)=1,sgn(\downarrow)=-1$ and $-\uparrow =\downarrow$.
The low energy, effective Hilbert space on an oxygen atom is four-dimensional.
On this Hilbert space it is easy to see that under the identification
\begin{eqnarray}
|\cdot ,\uparrow ,0,\cdot >= \left( \begin{array}{c}
                 1   \\
                 0   \\
                 0   \\
                 0   \\  \end{array} \right) \; , \;\;
|\cdot ,\downarrow ,0,\cdot >= \left( \begin{array}{c}
                 0   \\
                 1   \\
                 0   \\
                 0   \\  \end{array} \right) \; ,  \nonumber \\
|\cdot ,0 ,\uparrow,\cdot >= \left( \begin{array}{c}
                 0   \\
                 0   \\
                 1   \\
                 0   \\  \end{array} \right) \; , \;\;
|\cdot ,0 ,\downarrow,\cdot >= \left( \begin{array}{c}
                 0   \\
                 0   \\
                 0   \\
                 1   \\  \end{array} \right) \; , \label{2.12}
\end{eqnarray}
one finds, in the notation of the previous section,
\beq
Y^{\mu}({\vec x} ,m ) = \frac{\gamma^{\mu}({\vec x} ,m )
+i \rho ^{\mu} ({\vec x} ,m ) }{2} \; .               \label{2.13}
\eeq
The lattice gauge field is
\begin{eqnarray}
Y_{\alpha \; \beta}({\vec x} ,m ) &=& Y^{0}({\vec x} ,m )\; \delta_{\alpha \; \beta}
-i\vec Y({\vec x} ,m ) \; \cdot \; \vec \tau_{\alpha \; \beta} \nonumber \\
                       &=& \frac{1}{2} U_{\alpha \; \beta}({\vec x} ,m )
                       + \frac{i}{2} U^{5} _{\alpha \; \beta}({\vec x} ,m )  \label{2.14}
\end{eqnarray}
The effective Hamiltonian has a term generated by the process \rf{2.9}
\beq
H^{1}_{eff}=-\frac{t_{Cu-O} ^{2} }{A} \sum_{{\vec x} ,m} \sum_{\alpha ,\beta}
c^{\dag}_{{\vec x} ,\alpha} \;
Y_{\alpha ,\beta}({\vec x} ,m) \; c_{{\vec x} +a\vec{m} ,\beta} \; +\; h.c. ,  \label{2.15}
\eeq
which is an ${\rm SU}(2)$ gauge invariant hopping term.
 
The processes of cell excitation transport involving states \rf{2.8} have
not yet been considered. The
processes involving these these states have \rf{2.7} as intermediate
states. They are therefore included in the effective Hamiltonian by
introducing a term connecting $|\cdot ,\alpha ,0,\cdot >$ and
$|\cdot ,0 ,\alpha ,\cdot >$. This term can be read off from \rf{10.1} and
\rf{2.12} :
\beq
H^{2}_{eff}=t_{O} \sum_{{\vec x},m} \gamma^{0}({\vec x},m)\; .   \label{2.16}
\eeq
It breaks the ${\rm SU}(2)$ gauge invariance explicitly, by giving some
of the gauge spin waves a gap.
 
The operator \rf{1.17} is the same as
\beq
\vec G ({\vec x} )\; =\; \frac{1}{2} \sum_{\vec i \ep B_{{\vec x}} }
\sum_{\alpha \beta } \;
c^{\dag }_{\vec i ,\alpha} \; \vec \tau_{\alpha \beta }  \;
c_{\vec i ,\beta} \; ,  \label{2.161}
\eeq
which is just the total spin in a cell. The ``color" of the
${\rm SU}(2)$ gauge theory is simply spin.
This operator commutes with \rf{2.15}, but not \rf{2.16}.
 
Equation \rf{2.16} can also be viewed as the gauge invariant
Hamiltonian for an additional spin-zero field coupled to $Y$. The term
\beq
H^{2}_{eff}=t_{O} \sum_{{\vec x},m} \sum_{\alpha \beta \gamma}
\phi ^{\dag} _{\alpha \beta} ({\vec x})\;
Y_{\beta \gamma} ({\vec x},m)\; \phi _{\gamma \alpha } ({\vec x} +a\hat m)\; +h.c. , \label{2.17}
\eeq
where $\phi  ({\vec x} )$ is a unitary (c-number) $2 \times 2$
matrix, $ \phi ^{\dag} ({\vec x} ) \phi  ({\vec x} )=1$, provided the Gauss' law
operator is modified to
\begin{eqnarray}
\vec G ({\vec x})=\sum_{m} [\vec{{\tilde{\Sigma}}} ({\vec x} ,m)
+\vec \Sigma  ({\vec x} -a\hat m ,m) ]\; +\;
\frac{1}{2} \sum_{\alpha \beta }
c^{\dag} _{{\vec x} ,\alpha }
\vec{\tau}_{\alpha \beta}
c _{{\vec x} ,\alpha } \;   \nonumber \\
+\; \frac{1}{2} \sum_{\alpha \beta \gamma } \;
\vec{\tau}_{\alpha \beta}
\phi  _{\beta \gamma} ({\vec x})
\frac{\partial }{\partial \phi _{\gamma \alpha} ({\vec x})}\; -\frac{1}{2} \sum_{\alpha \beta \gamma } \;
\phi^{\dag} _{\alpha \beta} ({\vec x})
\vec{\tau}_{\beta \gamma}
\frac{\partial }{\partial {\phi ^{\dag }_{\gamma \alpha} ({{\vec x} })}}
\;, \; \label{2.18}
\end{eqnarray}
reduces to \rf{2.16} in a particular gauge (known as the ``unitary
gauge") in which $\phi _{\alpha \beta} ({\vec x} )=\delta_{\alpha \beta}$. The
field $\phi$ describes a chiral spin superfluid.
 
When the effect of electromagnetism is included the Hamiltonian must be
modified. Taking $A_{0} =0$ gauge, \rf{2.15} and \rf{2.17}
become, respectively :
\beq
H^{1}_{eff}=-\frac{t_{Cu-O} ^{2} }{A} \sum_{{\vec x} ,m} \sum_{\alpha ,\beta}
c^{\dag}_{{\vec x} ,\alpha} \;
Y_{\alpha \; \beta}({\vec x} ,m) \;
\exp\;[ ie \int_{{\vec x}}^{{\vec x} + a{\vec m} } A_{m} d x^{m}] \;\;
c_{{\vec x} +a{\vec m} ,\beta} \; +\; h.c. ,  \label{2.19}
\eeq
and
\begin{eqnarray}
H^{2}_{eff} &=& t_{O} \sum_{{\vec x},m} \sum_{\alpha \beta \gamma}
\phi ^{\dag} _{\alpha \beta} ({\vec x})\;
Y_{\beta \gamma} ({\vec x},m)\;                   \nonumber \\
            &\times &
\exp\;\left[i\left(1-\frac{2b}{a}\right) e \int_{{\vec x}}^{{\vec x} + a{\vec m} } A_{m} d x^{m}\right] \;\;
\phi _{\gamma \alpha } ({\vec x} +a\hat m)\; +h.c. . \label{2.20}
\end{eqnarray}
In \rf{2.20} the approximation was made that the vector potential $\vec A$
is smoothly varying (in the exact expression, the range of integration
in the Aharonov-Bohm phase factor is from ${\vec x} +b\vec m$ to ${\vec x}
+(a-b)\vec m$). The superfluid field has fractional charge
$(1-2b/a)\; e$. Even without proceeding further, it is clear that
this field already produces superconductivity. Cooper pairing of
cell excitations also occurs, making a total of two superfluid
condensates.
 
Thus far the part of the Hamiltonian depending only on the gauge field
has been ignored. Such a term will be generated by higher orders in the
hopping parameter expansion. The leading contribution is a plaquette
interaction
\beq
H^{3}_{eff}=(t_{Cu-O} /A )^{4}\sum_{{\vec x} } \sum_{ m \neq n}\; {\rm Tr}
Y({\vec x} ,m)Y({\vec x} +\hat m,n)Y({\vec x} +\hat n,m)^{\dagger}
Y({\vec x} ,n)^{\dagger}  \; +h.c. \; .          \label{2.21}
\eeq
If high frequency Fourier components of $c_{{\vec x} ,\alpha},\;
c^{\dag}_{{\vec x} ,\alpha}$
are integrated out, there is an additional contribution of the form \rf{2.21}.
The spin wave spectrum of \rf{2.21} will be studied elsewhere. The coefficient
of this term is extremely small in this perturbative analysis; but this
analysis is only meant to be a guide to obtaining $H_{eff}$. If
$t_{Cu-O}$ is larger than $A$, there is no reason to expect this term to
be small. In two space and one time dimension, a dynamical non-Abelian gauge
field coupled to a Higgs field will either confine or screen the sources
(which are holes at copper sites).
 
Holes in real high-temperature superconductors have a much lower
concentration than in this toy model. Only a small number of oxygen
atoms are actually doped, i.e. $2N \gg h$. The remaining
oxygen sites are not occupied
by holes in low-lying states. The system is described by the unextended
Hubbard model on the lattice of fig.2:
\begin{eqnarray}
H=-\sum_{<\vec i ,\vec j >} \sum_{\alpha }\; t_{\vec i ,\vec j } \;
c^{\dag} _{\vec i ,\alpha }c_{\vec j ,\alpha }\; +
\sum_{\vec i}\;U_{\vec i }\;
n_{\vec i,\uparrow}n_{\vec i,\downarrow}            \nonumber \\
+\mu [\sum_{i} (n_{\vec i,\uparrow}+n_{\vec i,\downarrow}) -h]\;.   \label{2.22}
\end{eqnarray}
Again there are
two hopping
parameters, $t_{{\vec x} + b\hat m ,\; {\vec x} +(a-b)\hat m}=-t_{O},\;
t_{{\vec x} ,\; {\vec x} \pm b\hat m}
=t_{Cu-O}$. There are two coulomb repulsion strengths, $U_{{\vec x}}
=U_{Cu}$ on copper sites and $U_{{\vec x} \pm b\hat m}=U_{O}$ on
oxygen sites. The issue is now whether the model \rf{2.1} is a good
description of the physics at distances of two or three lattice
spacings. I conjecture that undoped regions of a few lattice spacings
in diameter behave as cells connected by doped oxygen bonds, and that
\rf{2.1} arises as a real space renormalization of \rf{2.22}
 
\section{Confinement and Higgs Phases}
 
Ignoring the Higgs field $\phi$, the system will be in the confined phase.
That means that holes at the copper sites are confined into spin-singlet
pairs. Seperating a pair sufficiently far leads to the formation of a
``spino-electric" string between them. This string is a line of non-Abelian
electric flux. Its energy is proportional to its length (there is a
string tension). An operator which creates such a hole-string-hole
excitation on the ground state is
\beq
\beta(x,y;C)=c^{\dag}({\vec x}) \prod_{l\ep C} Y(l)\;
\tau ^{2} \; c^{\dag}(\vec z) \; ,                  \label{3.1}
\eeq
where $l$ is a link along the path $C$ between the holes at the
copper sites ${\vec x}$ and $\vec z$. This is an ${\rm SU}(2)$ baryon
creation operator. It is a bound state of two cell excitations. A
neutral meson-type excitation produced by
\beq
\m(x,y;C)=c^{\dag}({\vec x}) \prod_{l\ep C} Y(l)\;
d^{\dag }(\vec z) \; ,  \label{3.2}
\eeq
where $d^{\dag}(\vec z)$ creates an electron at a copper site,
can also exist, though it will have a much greater gap (because a single
electron on the copper atom has a very large energy). There are also
anti-baryon states containing two electrons. The string
can break only if new holes or electrons appear on copper sites, to
join to the new string ends. These particles must be pulled out of the Fermi
sea, at a large cost in energy (because if the string breaks, it is
inevitable that at least one electron is produced at a copper
site). Thus, unlike the situation in Q.C.D., fragmentation of strings
is rare. It is operators such as $\beta$ which
will condense in the ground state, leading to superconductivity.
 
Now suppose the Higgs field is coupled into the system. This field is a
unitary matrix which
transforms according to the fundamental representation of ${\rm SU}(2)$. It
breaks the effective gauge symmetry completely. Nonetheless a confined
phase is still conceivable. There are in fact two possible phases
for an ${\rm SU}(N)$ gauge theory in two space and one time dimension.
The other phase, in which the Higgs field screens adjoint sources, is
called the Higgs phase \cite{'tHooft}. This
phase also has a gap. Fradkin and Shenker \cite{FS}
showed that in this situation, the phases are one and the same. In the
Euclidean lattice formulation, the phase boundary terminates in a
critical point, beyond which
the phases are connected. The physical reason is that the basic excitations
in the confined phase and the Higgs phase can be made by acting
with the same operators on the ground state. A pair of holes is no
longer bound by a string, but is instead screened by Higgs quasiparticles.
The operator which makes these quasiparticles on the ground state is, in the
unitary gauge, the same operator which created the spino-electric string
in the confined phase. Therefore pairing and superconductivity
will still occur. The size of the pair will now be determined by
the screening length instead of the string tension.
 
\section{Conclusions}
 
By including the effect of local lattice distortions in $CuO_{2}$
layers, a new kind of Hubbard model has been proposed, with two sites
on the links of the lattice. By integrating out high frequency modes in
strong coupling perturbation theory a lattice gauge magnet was obtained. The
resulting theory confines or screens the spin of hole quasiparticles
at the intersection points (copper atoms), resulting in pairing and
superconductivity. Because of the role of oxygen nuclear motion,
at least a weak oxygen isotope effect should result. There is an second
charge condensate, corresponding to the Goldstone mode of the fractionally
charged, spinning Higgs field.
 
This theory of superconductivity
has an appealing feature. It is an attempt to
describe the physics at all relevant wavelengths, from
the interatomic spacing to the macroscopic effective Hamiltonian. Most
of the field theoretic ideas are in accord with long prevailing
conventional wisdom.
 
Although perturbation theory was used to obtain this result, it is only
good quantitatively for very small hopping parameters $t_{Cu-O},
t_{O}$. Even though $t_{O}$ is much larger than the energy scale
$A$, a gauge theory of the sort derived here should still
describe the effective low frequency dynamics, provided $t_{O}/A$
is below a certain critical value. At this value there is a phase
transition to the $t-J$ model.
 
The mechanism proposed is not fundamentally
two-dimensional. Three-dimension- al
Hubbard models can also be described by non-Abelian gauge theories. In
three dimensions we expect that the phase transition, to a phase
in which the holes are unbound, occurs at a lower value of $T_{c}$. Perhaps
three-dimensional superconducting bismuth oxide
materials such as $Ba_{1-x} Pb_{x} Bi_{x} O_{3} $ and
$Ba_{1-x} K_{x} Bi O_{3-y} $ \cite{Bis} are also described by
the picture presented here.
 
\section{Acknowledgements}
I thank T. K. Lee and Liam Coffey for educating me about the
phenomenology of high temperature superconductors; though they aren't
responsible for any remaining misconceptions on my part. Conversations
with Michael El-Batanouny and Gergei Zemanyi several years ago
had a profound effect on the picture presented here. I
thank Holger Bech Nielsen for discussions about the
origin of gauge invariance and the nature of the excitations. I am
grateful to A.A. Nersesyan for encouragement. Philip
Stamp pointed out a fundamental error in an earlier attempt
to apply the idea of gauge magnets to superconductivity. This work also
benefited from discussions with
Lay Nam Chang, Alan Luther, Poul
Olesen, Daniel Rohrlich, Gordon Semenoff, Mike Stone
and Chia Tze. Finally I thank the Niels Bohr Institute staff
for their hospitality.

\vfill
\newpage

\section{Figure Captions}
\begin{itemize}

\item Figure 1:\begin{description} 
               \item[]a) Oxygen displacement for a hole in the oxygen
                         orbital near copper atom A. In this example
                         the oxygen atom is attracted by the hole.
               \item[]b) Oxygen displacement for a hole in the oxygen
                         orbital near copper atom B.
               \item[]c) Effective double well potential.
               \end{description}  

\item Figure 2: The lattice of the Hubbard model. The dotted lines enclose
                a cell.

\item Figure 3: A low-lying configuration of holes for hole number equal
                to $2N$. Note that the hole positions resemble those of
                of hydrogen ions in two-dimensional ice models.

\item Figure 4: A low-lying configuration of holes for hole number slightly
                greater than $2N$. Excited cells are shaded.

\item Figure 5: The energy parameters $A$, $t_{Cu-O}$, $-t_{O}$. On the left 
                side of the figure, $t_{O}$ is in the regime where the
                hopping parameter expansion is valid. This parameter
                increases from the left to the right side of the 
                figure. Eventually there is a phase transition to where
                the physics is described by the $t-J$ model.
                
\item Figure 6:\begin{description} 

               \item[]a) An excitation in the cell at the left. 

               \item[]b) Another example of such an excitation.

               \end{description}  

\end{itemize}

\begin{center}
\begin{figure}
\centering
\includegraphics[width=7in]{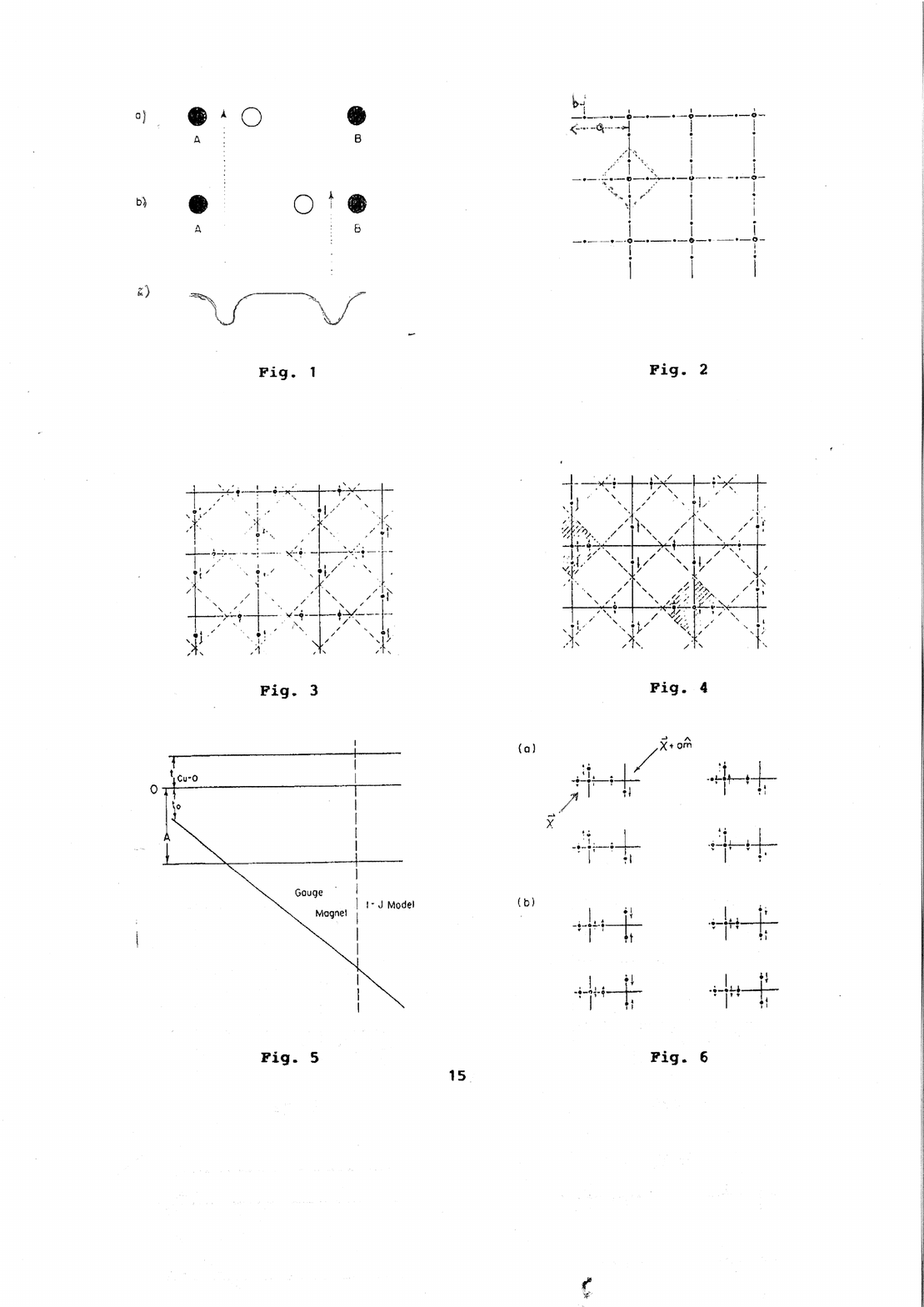} 
\end{figure}
\end{center}

\end{document}